\documentstyle[12pt]{article}
\begin{document}
\newcommand{\be}{\begin{equation}}
\newcommand{\ee}{\end{equation}}
\newcommand{\bea}{\begin{eqnarray}}
\newcommand{\eea}{\end{eqnarray}}
\newcommand{\sptwo}{1.4}
\newcommand{\doublespace}{\edef\baselinestretch{\sptwo}\Large
\normalsize}

\vspace{0.5in}
\begin{center}
{\large\bf On the Holomorphic Structure of a Low Energy 
Supersymmetric Wilson Effective Action}
\end{center}
\vspace{0.5in}
\begin{center}
{\bf T.E. Clark and S.T. Love}\\
{\it Department of Physics\\ 
Purdue University\\
West Lafayette, IN 47907-1396}
\end{center}
\vspace{1.0in}
\begin{center}
{\bf Abstract}
\end{center}

The Wilson (exact) renormalization group equations are used to 
determine the evolution of a general low energy 
N=1 supersymmetric action containing a U(1) gauge vector 
multiplet and a neutral chiral multiplet.  The effective theory 
evolves towards satisfying a fixed relation where the K\"ahler 
potential and effective gauge coupling are obtained 
from a N=2 supersymmetric holomorphic prepotential. 
\pagebreak

\doublespace

As a consequence 
of  the non-renormalization properties of their 
radiative corrections, the low energy structure of many 
supersymmetric models can be exactly 
determined.  This attribute was first displayed in calculations 
of the superpotential in purely 
perturbative models \cite{FL}\cite{GRS}.  
Subsequently, it was recognized that by combining 
various symmetries with the holomorphic 
dependence on the fields and parameters of a model, the 
superpotential can also be completely (non-perturbatively) 
obtained even in the 
framework of strongly interacting gauge theories \cite{Seib}.  
For N=2 supersymmetric theories, the 
low energy action is given in terms
of a single holomorphic prepotential.  As in the N=1 superpotential case, 
this prepotential is exactly determined using the symmetries, 
holomorphicity and duality properties of the model in question.  
Seiberg and Witten \cite{SW} secured the form of the 
prepotential (and hence the K\"ahler metric on the quantum moduli space) in a 
N=2, SU(2) strongly interacting gauge theory which is spontaneously broken to 
a low energy N=2, U(1) gauge theory.  
In terms of N=1 superfields, this low energy action has the form
\bea
       \label{N=2action}
      \Gamma_{N=2} [\phi, \bar\phi, V] &=& \frac{1}{8\pi i}
      \left(\int dV \left[ {\cal F}_\varphi(\varphi)\bar\varphi-
      \bar{\cal F}_{\bar\varphi}(\bar\varphi)\varphi \right]\right. \nonumber\\
       & & \left.+\frac{1}{2} \int dS
      {\cal F}_{\varphi\varphi}(\varphi)W^{\alpha}W_\alpha -
      \frac{1}{2} \int d\bar S \bar{\cal F}_{\bar\varphi\bar
      \varphi}(\bar{\varphi})
      \bar W_{\dot\alpha}\bar W^{\dot\alpha} \right) ,\nonumber\\
&&
\eea
where $(\bar\varphi)$ $\varphi$ is a neutral (anti-) 
chiral superfield, the N=2 partner of the 
abelian gauge field $V$. Here the subscripts denote differentiation with 
respect to that field so that, 
for example, ${\cal F}_\varphi=\frac{\partial{\cal F}}{\partial\varphi}$. 
The chiral 
field strength is defined by $W_\alpha =-\frac{1}{4}\bar 
D\bar D (e^{-V}D_\alpha e^{V})= -\frac{1}{4}\bar D\bar D D_\alpha V$, 
with a similar definition, 
$\bar W_{\dot\alpha}= -\frac{1}{4}DD\bar D_{\dot\alpha}V$, 
for the anti-chiral field strength.  
The holomorphic prepotential ${\cal F}(\varphi)$ then 
determines the K\"ahler potential $K(\varphi,\bar{\varphi})=\frac{1}{4\pi}
{\rm Im}[{\cal F}_\varphi 
(\varphi) \bar\varphi]$ and hence the K\"ahler 
metric 
\be
g(\varphi,\bar\varphi)=K_{\varphi\bar{\varphi}}(\varphi,\bar{\varphi})=
\frac{1}{4\pi}{\rm Im} {~\cal F}_{\varphi\varphi}.
\ee
Seiberg and Witten constructed $g(\varphi,\bar\varphi)$ 
for a strongly interacting N=2 SU(2) 
Yang-Mills theory.  Analogous results have also been obtained in other 
models which include matter fields as well as different gauge groups \cite{X}.

A general low energy effective action for an arbitrary 
N=1 supersymmetric theory 
containing these same fields can be written as
\bea
     \label{action}
    \Gamma [\phi, \bar\phi, V] &=& \int dV K(\varphi, \bar\varphi) 
     +\frac{1}{2} \int dS
     {f}(\varphi)W^{\alpha}W_\alpha +\frac{1}{2} 
     \int d\bar S \bar{f}(\bar{\varphi})
     \bar W_{\dot\alpha}\bar W^{\dot\alpha}  \nonumber\\
     & & +\kappa \int dV V +\int dS P(\varphi) +\int d\bar 
     S \bar P(\bar\varphi) .
     \eea
Here the K\"ahler potential, $K(\varphi,\bar\varphi)$, 
the effective gauge couplings, $f(\varphi)$ and 
$\bar f(\bar\varphi)$, and the superpotentials, 
$P(\varphi)$ and $\bar P(\bar\varphi)$, are arbitrary, in general, 
unrelated functions of their respective 
arguments while the linear in $V$ Fayet-Iliopolous term 
has an arbitrary, but field independent, 
coupling constant $\kappa$.  These functions and couplings evolve under 
renormalization group transformations operating at the 
scale at which the effective theory is defined.  
As is well known, the superpotential does not evolve 
independently but only according to the anomalous dimension of the 
chiral fields.  In this topologically trivial 
model, if it initially vanishes, the superpotential remains zero.  
Likewise, the Fayet-Iliopolous term has no independent 
additive radiative corrections. Consequently its zero value is 
stable as the system evolves.  Furthermore, after 
choosing these terms to vanish, we shall use the low energy truncation of the 
Wilson renormalization group equations to establish that 
the general N=1 supersymmetric effective action (\ref{action}) 
evolves toward satisfying the N=2 relation between the 
K\"ahler potential and the effective gauge couplings 
given by $g(\varphi,\bar\varphi)=f(\varphi)+\bar f(\bar\varphi)$ 
as the theory flows toward the infrared stable trivial fixed point. 
Such a renormalization group behavior in N=1 SUSY theories is 
reminiscent of previous studies\cite{LS} of various different 
non-abelian gauge models which exhibit a manifold of 
infrared attractive stable N=2 supersymmetric non-trivial fixed points towards 
which the theory evolves. The present abelian N=1 
model exhibits a similar behavior except in this case, the N=2 fixed point 
is the trivial one.

The Wilson renormalization group equation (WRGE) describes the 
change of the (Wilson) effective action as the quantum mechanical 
effects of the degrees of freedom with momentum differentially below 
the running cutoff $\Lambda (t) =e^{-t} \Lambda$ are taken into 
account \cite{W}\cite{HW}.  It is secured by demanding that correlation 
functions remain unchanged as the degrees of freedom in the differential 
shell are integrated out. As such, the action satisfying the WRGE 
can be used to describe the physics on all scales below $\Lambda (t)$. 
Making a momentum expansion of the general effective action and 
retaining all terms containing two or less space-time derivatives, 
then the WRGE describes the flow of the low energy action (\ref{action}).  
Although such an action is a trunction of the space of all possible actions, 
its self-radiative corrections will be consistently 
determined by this method, independent of the strength of the interaction.  
Integrating over vector and chiral superfields with momentum 
between $e^{-t}\Lambda (t)$ and $\Lambda (t)$ while re-scaling all 
dimensionful parameters by $\Lambda (t)$ and all fields according to 
their anomalous dimensionality, we secure the Wilson renormalization 
group equations for the action (\ref{action}).  These radiative corrections 
arise only from the one loop one particle irreducible diagrams with all 
internal lines having momentum at the cutoff \cite{CHL}.  

The superpotential receives no radiative corrections and its Wilson 
renormalization group equation is simply given by
\bea
      {\partial
      P\over \partial t} &=& 3P-(1-\gamma ) \varphi P_\varphi \nonumber\\
       {\partial \bar P\over \partial t} &=& 3\bar P-(1-\gamma ) 
      \bar\varphi \bar P_{\bar\varphi} .
\eea
Here $\gamma$ is the anomalous dimension for the (anti-) 
chiral superfield.  
It follows that if the superpotential vanishes at $t=0$, it will remain 
zero for all $t>0$.  This is simply the Wilson renormalization group equation 
verification of the non-renormalization 
theorem for the superpotential in this model.  Likewise in this model, the 
Fayet-Iliopoulous term receives no radiative corrections.  
Thus if initially zero, it also remains zero.  

On the other hand, the K\"ahler potential does, in general, 
receive radiative corrections.  These can be computed by evaluating the graphs 
in which both chiral fields and the 
abelian gauge fields propagate in the loop with external legs carrying 
zero momentum. Once again defining the 
K\"ahler metric as  $g(\varphi,\bar\varphi)=K_{\varphi\bar\varphi}(\varphi,
\bar\varphi)$, we find the Wilson renormalization group equation for $g$ 
takes the form
\bea
\label{wrge1}
     {\partial g\over \partial t} &=& 2\gamma g -(1-\gamma)
     \varphi g_\varphi -(1-\gamma)\bar
     \varphi g_{\bar\varphi} \nonumber\\
     & & +{1\over 8\pi^2}\left({g_\varphi g_{\bar\varphi}\over g^2}-
      {g_{\varphi\bar\varphi}\over  g}-{f_\varphi
     \bar f_{\bar\varphi}\over (f+\bar f)^2}\right) .
\eea
In obtaining the above result, we have employed 
an $R_\xi$ gauge so that the action is augmented by the term
\be
\Gamma_{\xi} [V] = \frac{\xi}{8} \int dV  (DDV)(\bar D\bar D V).
\ee
Note that the gauge 
fixing parameter $\xi$ receives no radiative corrections in this 
simple abelian model.  As such, it does not contribute to the 
K\"ahler potential radiative
corrections.  
Since the fields have been rescaled at each $t$ during the renormalization 
flow according to their full scaling  
dimensions, the chiral field anomalous 
dimension can be extracted by evaluating equation (\ref{wrge1}) 
at $\varphi=\bar\varphi=0$ where 
$g\vert_{\varphi = \bar\varphi= 0}={1}$ and 
$f\vert_{\varphi =0}=\bar f_{\bar\varphi =0}={1\over 2}$.  
So doing, we find 
\be
2\gamma = {1\over 8\pi^2}\left[ g_{\varphi\bar\varphi} 
-g_\varphi g_{\bar\varphi} +f_\varphi \bar f_{\bar\varphi} 
\right]\vert_{\varphi = \bar\varphi=0} .
\ee
Using Eq. 
(\ref{wrge1}), the K\"ahler potential is seen to satisfy the Wilson 
renormalization group equation
\be
{\partial K\over \partial t} = 
2K-(1-\gamma)\varphi K_\varphi 
-(1-\gamma)\bar\varphi K_{\bar\varphi}-{1\over 8\pi^2}
\ln({K_{\varphi\bar\varphi}\over f+\bar f}).
\ee
Note that the K\"ahler potential is only determined up to additive 
holomorphic chiral and anti-chiral functions so that 
$K$ and $K+F(\varphi) +\bar F(\bar\varphi)$ give rise to the same metric 
and hence produce the same physics  \cite{BW}\cite{CL} .

Next turning to the running of 
the holomorphic effective gauge couplings $f$ and $\bar f$, 
here one again finds that the radiative loop corrections identically vanish so 
that they also 
evolve only according to the anomalous dimensionality of the fields. 
The resultant Wilson renormalization group equations take the simple form
\bea
\label{wrge2}
{\partial f\over \partial t} &=& 2\gamma_V  f -(1-\gamma)\varphi f_\varphi 
\nonumber\\
{\partial \bar f\over \partial t} &=& 2\gamma_V  \bar f -(1-\gamma)\bar\varphi 
\bar f_{\bar\varphi} ,
\eea
where the photon anomalous dimension is denoted as $\gamma_V$. 
By again evaluating these equations at zero fields, 
we immediately glean that $\gamma_V =0$.  
Hence the scaling of effective gauge couplings is secured as 
\bea
f(\varphi , t) &=& f(e^{\int_0^t dt (\gamma-1)}\varphi,0)=  
\tau (e^{\int_0^t dt (\gamma-1)}\varphi) \nonumber\\ 
\bar f(\bar\varphi , t) &=& \bar f(e^{\int_0^t dt (\gamma-1)}\bar\varphi,0)=
\bar\tau (e^{\int_0^t dt (\gamma-1)}\bar\varphi) , 
\eea
where $\tau (\varphi)$ and 
$\bar\tau (\bar\varphi)$ are the initial holomorphic 
effective gauge couplings.  

Now consider Eq. (\ref{wrge1}) satisfied by the K\"ahler metric. 
Note that when  
\be
g(\varphi,\bar\varphi)= f(\varphi)+\bar f(\bar\varphi),
\ee 
so that $g_\varphi =f_\varphi$ 
and $f_{\bar\varphi}=\bar f_{\bar\varphi}$ while $g_
{\varphi\bar\varphi}=0$, then the loop corrections identically vanish.  
As such this can be viewed as a fixed relation of the Wilson renormalization 
group equation. Moreover, when this functional relation is satisfied, 
it also follows that $\gamma =0$. We thus secure the metric 
form 
\be
g(\varphi,\bar\varphi)= f(\varphi)+\bar f(\bar\varphi)=\tau (e^{-t}\varphi)
+\bar\tau (e^{-t}\bar\varphi).
\ee  

Note that an immediate consequence of this result is that if the action is 
initially N=2 supersymmetric, then  
it remains N=2 supersymmetric with a  
holomorphic prepotential ${\cal F}$ whose functional form is fixed and 
which evolves into the infrared 
according to the above naive dimensional scaling of the fields.  
For this N=2 abelian model, the form of the 
next higher order term ($p^4$) in the momentum 
expansion of the action has also been 
investigated \cite{H}. Its structure requires the introduction of an 
additional real analytic N=2 supersymmetric function of the fields.  
On the other hand, an explicit superspace one-loop perturbative calculation 
\cite{GRvU}\cite{dWGR}\cite{PW} 
for an N=2 non-abelian gauge theory which also contains 
a superpotential exhibits a breakdown of the manifest N=2 
invariance and the K\"ahler potential cannot be written in terms of a single 
holomorphic function.

To establish that the N=2 supersymmetric relation $g= (f+\bar f)$ is an 
attractive infrared stable fixed condition, 
the Wilson equations (\ref{wrge1}) and (\ref{wrge2}) are 
expanded for small field values. So doing, it follows that the ratios    
\bea
r &\equiv & {g_{\varphi}|_{\varphi=\bar{\varphi}=0} 
\over f_{\varphi}|_{\varphi=0}}\nonumber\\
\bar r &\equiv & {\bar g_{\bar{\varphi}}|_{\varphi=\bar{\varphi}=0}\over 
\bar f_{\bar{\varphi}}|_{\bar{\varphi}=0} },
\eea
satisfy the renormalization group equations
\bea
\label{RGEL}
{1\over r}{d r\over d t} &=& -{1\over 8\pi^2}
[f_{\varphi} \bar f_{\bar{\varphi}}]|_{\varphi=\bar{\varphi}=0} 
\left[ (r\bar r -1) +2(r\bar r - {1\over r})\right]\nonumber\\
{1\over \bar r}{d \bar r\over d t} 
&=& -{1\over 8\pi^2}
[f_{\varphi} \bar f_{\bar{\varphi}}]|_{\varphi=\bar{\varphi}=0} 
\left[ (r\bar r -1) +2(r\bar r - {1\over \bar r})\right] .
\eea
The N=2 relation $g=f+\bar{f}$ requires that 
$g_{\varphi}|_{\varphi=\bar{\varphi}=0}=f_{\varphi}|_{\varphi=0}$ and 
$g_{\bar{\varphi}}|_{\varphi=\bar{\varphi}=0}=
\bar{f}_{\bar{\varphi}}|_{\bar{\varphi}=0}$. These in turn correspond to the 
values $r=\bar{r}=1$ which are clearly fixed points of Eq. (\ref{RGEL}). 
Considering small deviations from the fixed point values, so that 
$r=1 +\epsilon$ and $\bar r =1+\bar\epsilon$, we find that
\bea
{d r\over d t} &=& -{1\over 8\pi^2}
[f_{\varphi} \bar f_{\bar{\varphi}}]|_{\varphi=\bar{\varphi}=0} 
\left[ 5\epsilon +3\bar\epsilon\right]\nonumber\\
{d \bar r\over d t} &=& -{1\over 8\pi^2}
[f_{\varphi} \bar f_{\bar{\varphi}}]|_{\varphi=\bar{\varphi}=0} 
\left[ 5\bar\epsilon +3\epsilon\right] .
\eea
Hence, the N=2 fixed point  $r=\bar r=1$ is indeed attractive.  The 
N=1 supersymmetric theory evolves 
towards the N=2 supersymmetric theory characterized by the relation 
$g=(f+\bar f)$.

In summary, we have examined the structure of the 
low energy Wilson effective action containing abelian vector and neutral chiral 
superfields as the theory evolves into the infrared.  
The self-radiative corrections of this N=1 
supersymmetric theory drive the low energy 
degrees of freedom to a N=2 symmetric 
theory. If the action is initially N=2 supersymmetric, then  
it remains N=2 supersymmetric with a holomorphic prepotential ${\cal F}$ 
whose functional form is fixed and which evolves into the infrared 
according to the naive dimensional scaling of the fields.  
\smallskip

This work was supported in part by the U.S. Department 
of Energy under grant DE-FG02-91ER40681 (Task B).
\pagebreak

\newpage
\end{document}